\documentclass[sigconf]{acmart}

\AtBeginDocument{%
  \providecommand\BibTeX{{%
    \normalfont B\kern-0.5em{\scshape i\kern-0.25em b}\kern-0.8em\TeX}}}

\copyrightyear{2024} 
\acmYear{2024} 
\setcopyright{acmlicensed}\acmConference[CAIN 2024]{Conference on AI Engineering Software Engineering for AI}{April 14--15, 2024}{Lisbon, Portugal}
\acmBooktitle{Conference on AI Engineering Software Engineering for AI (CAIN 2024), April 14--15, 2024, Lisbon, Portugal}
\acmDOI{10.1145/3644815.3644954}
\acmISBN{979-8-4007-0591-5/24/04}

\begin{document}

\title{What About the Data? A Mapping Study on Data Engineering for AI Systems}


\author{Petra Heck}
\email{p.heck@fontys.nl}
\orcid{0000-0002-9378-3213}
\affiliation{%
  \institution{Fontys University of Applied Sciences}
  \city{Eindhoven}
  \country{Netherlands}
}
\renewcommand{\shortauthors}{Heck}

\begin{abstract}
AI systems cannot exist without data. Now that AI models (data science and AI) have matured and are readily available to apply in practice, most organizations struggle with the data infrastructure to do so. There is a growing need for data engineers that know how to prepare data for AI systems or that can setup enterprise-wide data architectures for analytical projects. But until now, the data engineering part of AI engineering has not been getting much attention, in favor of discussing the modeling part. In this paper we aim to change this by perform a mapping study on data engineering for AI systems, i.e., \textit{AI data engineering}. We found 25 relevant papers between January 2019 and June 2023, explaining AI data engineering activities. We identify which life cycle phases are covered, which technical solutions or architectures are proposed and which lessons learned are presented. We end by an overall discussion of the papers with implications for practitioners and researchers. This paper creates an overview of the body of knowledge on data engineering for AI. This overview is useful for practitioners to identify solutions and best practices as well as for researchers to identify gaps.     
\end{abstract}




\maketitle

\section{Introduction}
\begin{figure}[h]
  \centering
  \includegraphics[width=\linewidth]{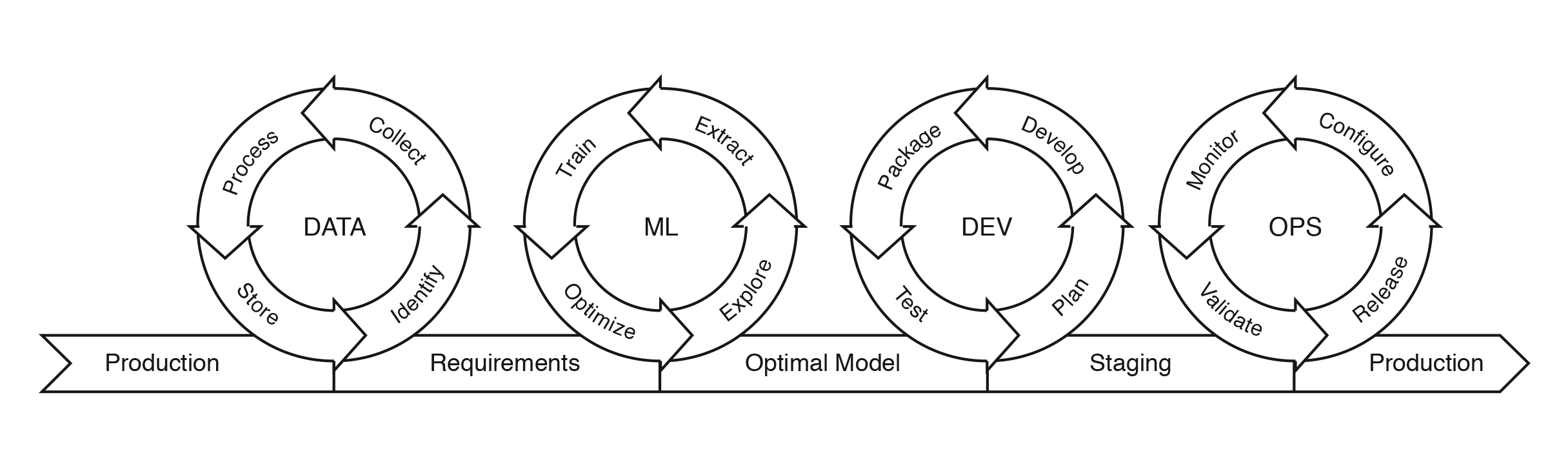}
  \caption{The AI engineering life cycle ~\cite{Farah}}
  \Description{The AI engineering life cycle ~\cite{Farah}}
  \label{fig:phasesAIEng}
\end{figure}

AI systems cannot exist without data~\cite{groger}. To develop an AI system, data needs to be collected and prepared to train the AI model, see the DATA cycle in Figure \ref{fig:phasesAIEng}. But also when the model is in production, production data needs to be prepared to send it to the model and get back predictions. This data can be structured tabular data or unstructured data such as images, sound, text or video. In 2021, Andrew Ng coined the term ``data-centric AI'' for ``the discipline of systematically engineering the data used to build an AI system''\footnote{\url{https://datacentricai.org/}}, i.e., \textit{AI data engineering}.  

\begin{figure}[h]
  \centering
  \includegraphics[width=\linewidth]{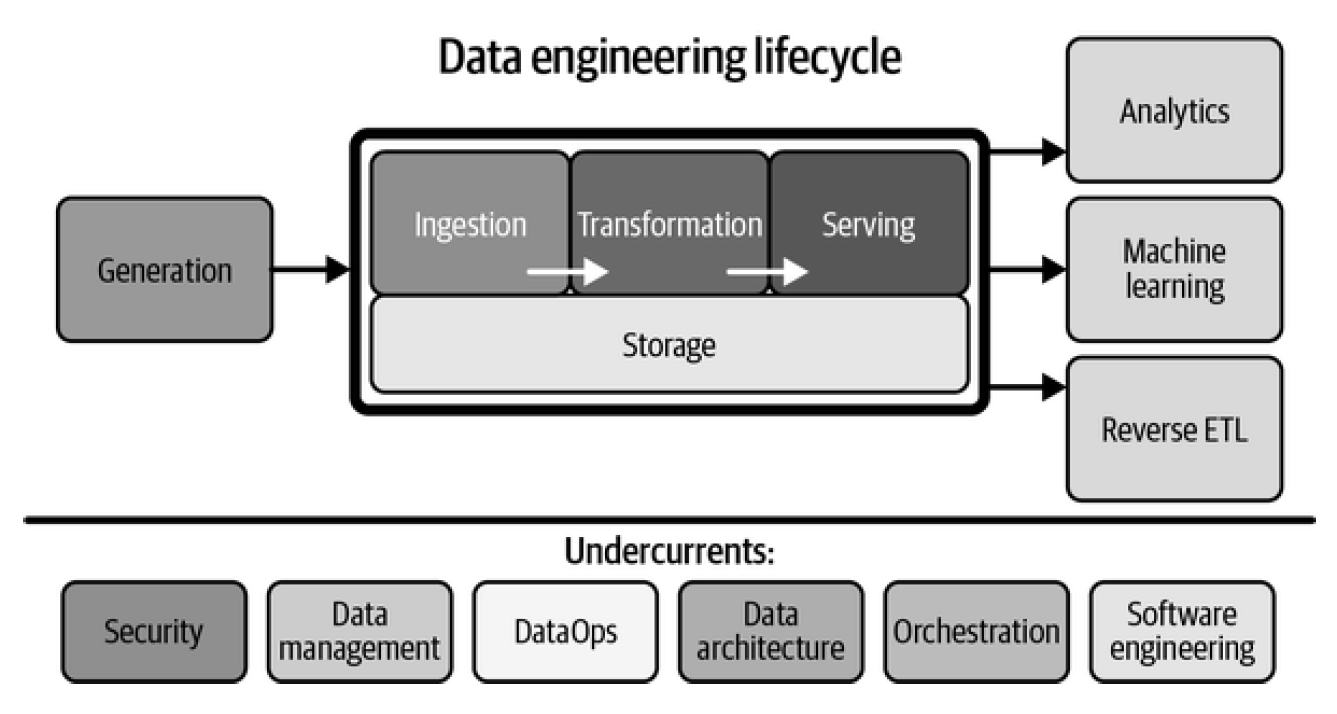}
  \caption{The data engineering life cycle ~\cite{Reis}}
  \Description{The data engineering life cycle ~\cite{Reis}}
  \label{fig:DE}
\end{figure}

Reis and Housley~\cite{Reis} define data engineering as ``the development, implementation, and maintenance, of systems and processes that take in raw data and produce high-quality, consistent information that supports downstream use cases, such as analysis and machine learning. Data engineering is the intersection of security, data management, DataOps, data architecture, orchestration and software engineering.'' Figure \ref{fig:DE} shows the data engineering life cycle, consisting of Ingestion, Transformation, Serving and Storage. The definition of Reis and Housley is very broad, not just focused on one AI engineering project as is the life cycle in Figure \ref{fig:phasesAIEng}, and also has a clear link to machine learning (thus AI). 

In the projects we do with industry we see that now AI models have matured and are readily available to apply in practice, most organizations struggle with the data infrastructure to do so. This can be on a project level: How to retrain this pre-trained model with our own small dataset? How to create synthetic data? How to merge different data sources? How to store and version data? How to automate data pre-processing? But also on organization level: How to integrate all available data from different systems into a central ``store''? How to protect sensitive data? How to deal with privacy issues? How to clean data? How to design an organization-wide data architecture that is fit for future AI developments? More and more organizations directly ask for data engineers as the main driver for their AI initiatives~\cite{heck2022}. They struggle to find these employees, because software engineers have not been trained on "big data" or data architectures specifically and data scientists do not have the required software engineering skills. Both disciplines rather focus on the more attractive model development part of AI engineering. 

Jarrahi et al.~\cite{Jarrahi} also stress the importance of data-centric AI (DCAI). They state that ``the nature of `data work' itself is not necessarily new. However, over the years, the actual data work in AI projects comes mostly from individual initiatives, and/or from piecemeal and ad hoc efforts. A lack of attention to data excellence and quality of data has resulted in underwhelming outcomes for AI systems, particularly those deployed in high-stake domains such as medical diagnosis. DCAI magnifies the role of data throughout the AI life cycle and stretches its lifespan beyond the so-called `preprocessing step' in model-centric AI.''

One could argue that to account for true DCAI the AI engineering life cycle should also be extended with undercurrents such as ``data architecture'' and ``DataOps'', analogue to Figure \ref{fig:DE}. These are data-related activities that are not just relevant for one single AI engineering project, but for the entire organization, or for all projects being executed, or for all AI systems being maintained. This is in fact also what we see in the projects we do with industry. They know how to quickly get data for one machine learning experiment, but not how to set up a data architecture for enterprise-scale AI engineering.  

This paper sets out to answer the question \textit{``How to do data engineering for AI systems?''} by means of a mapping study. For this mapping study we formulated the following research questions: 
\begin{itemize}
\item RQ1: Which data and AI engineering lifecycle phases are covered?
\item RQ2: Which technical solutions (tools/frameworks/platforms) for AI data engineering are proposed?
\item RQ3: Which architectures for AI data engineering are proposed?
\item RQ4: What are lesson learned on AI data engineering?
\end{itemize}

The mapping study identified 25 papers that explain data engineering activities, tools, frameworks or architectures. By categorizing them and summarizing their solutions and lesson learned, the paper creates an overview of the body of knowledge on data engineering for AI. This overview is useful for practitioners to identify solutions as well as for researchers to identify gaps.   

The remainder of the paper is organized as follows. Section \ref{sec_related} provides related work on data engineering for AI systems. Section \ref{sec:method} explains the method used for the mapping study and how the 25 resulting papers were selected. Section \ref{sec:data} classifies the 25 papers according to their meta-data, the type of solution they discuss and the scope of the data engineering activities explained. Section \ref{RQ} analyzes each of the four subquestions (RQ1 till RQ4). Section \ref{sec_discuss} discusses the overall research question and implications for practitioners and researchers. The paper concludes with main findings and future work.  

\section{Related Work} \label{sec_related}
This section describes related work on data engineering within AI engineering research.

Amershi et al.~\cite{Amershi2019} is one of the first AI engineering case studies to appear. The paper includes a machine learning workflow including data-oriented steps. It describes data engineering challenges at Microsoft but not really explains solutions. The paper shows that already in 2019, data management and data discovery on a project-level was a challenge for AI engineers. The case study does not describe data engineering challenges on enterprise level.

Serban et al.~\cite{Serban} identified a set of best practices for AI engineering from existing literature, including five best practices on data: 
\begin{enumerate}
    \item Use sanity checks for all external data sources
    \item Check that input data is complete, balanced and well distributed
    \item Write reusable scripts for data cleaning and merging
    \item Ensure data labelling is performed in a strictly controlled process
    \item Make data sets available on shared infrastructure (private or public)
\end{enumerate}
More information on these best practices (and how to implement them) might be taken from the papers that list them. Since Serban et al. is a meta-research it is not included in our mapping study. 

Several authors ~\cite{bosch2021, Ozkaya, Fischer, Nahar, Shivashankar, Heyn, sambasivan, Munappy} describe data challenges for AI systems, without offering explicit solutions. Each of those papers does however stress the importance of data engineering for AI systems. Bosch et al.~\cite{bosch2021} present DataOps as part of the AI engineering research agenda as a ``a significant opportunity to reduce ... overhead by generating, distributing and storing data smarter in the development process''. Sambasivan et al.~\cite{sambasivan} state that ``Data quality carries an elevated significance in high-stakes AI due to its heightened downstream impact, impacting predictions like cancer detection, wildlife poaching, and loan allocations. Paradoxically, data is the most under-valued and de-glamorised aspect of AI.''. This paper complements previous work with an overview of actionable solutions. 

\section{Paper Selection}\label{sec:method}
To select papers that answer our research questions on data engineering for AI we used the process described by Kitchenham and Charters ~~\cite{Kitchenham}: 
\begin{enumerate}
    \item Define inclusion and exclusion criteria
    \item Design query string
    \item Identify databases and other sources to search
    \item Select relevant papers based on title and abstract
    \item Select relevant papers based on full text
    \item Extend result set based on citations
    \item Classify resulting papers
\end{enumerate}
As a final step, we coded ~\cite{Saldana2011} the resulting papers to answer the research questions. 
The following paragraphs describe each of the steps in detail. The classification step is described in Section~\ref{sec:data}.

\subsection{Inclusion and Exclusion Criteria}
We used four \emph{inclusion} criteria. Three of them are on metadata of the paper: the paper should be in English, peer-reviewed and from 2019 or later. We choose 2019 because the AI engineering publications are all from that year or later, e.g., the seminal Microsoft case study by Amershi et al.~\cite{amershiCorrected}. The fourth inclusion criterion is on the content of the paper. The paper should explain data engineering activities, best practices, tools, frameworks or data architectures in the context of AI engineering (building production-ready AI systems).
We \emph{exclude} books, theses (bachelor, master and PhD.) and meta-research such as systematic literature reviews or mapping studies. Furthermore we excluded papers that only mention challenges, without providing solutions.  

\subsection{Query String}
As we are looking for data engineering for AI it makes sense to include both these terms in the query. We choose to use the more specific term ``AI engineering'' since we are interested in the software engineering view on AI. We also included ``data requirements'' and ``data architecture'' to surface papers that describe data artefacts (and thus possibly data activities) without mentioning ``data engineering''.
\begin{quote}
\textbf{(data engineering || data requirements || data architecture) \& AI engineering} \\
\end{quote}

\subsection{Source Selection and Query Execution}
We selected five digital databases which index software engineering venues plus Google Scholar. Furthermore we chose to try out Elicit, an AI-powered research assistant. Elicit uses language models to extract data from and summarize research papers. A search in Elicit is based on a question that you are trying to answer, not on a query string. We directly asked Elicit to come up with papers that answer the question ``How to do data engineering for AI engineering''. Elicit generates papers in batches, and we stopped when a newly generated batch did not contain relevant titles anymore.  
We executed the query string on each digital database (June 2023), resulting in 259 unique items in total\footnote{Dataset available online at \url{https://dx.doi.org/10.13140/RG.2.2.17748.78725}}:
\begin{itemize}
\item{Google Scholar (\url{scholar.google.com}) 182 items}
\item{IEEE Xplore (\url{ieeexplore.ieee.org}) 5 items, 0 unique items}
\item{ACM Digital Library (\url{dl.acm.org}) 8 items, 1 unique item}
\item{Scopus (\url{scopus.com}) 3 items, 0 unique items}
\item{DBLP (\url{dblp.org}) 16 items, 13 unique items}
\item{ScienceDirect (\url{sciencedirect.com}) 7 items, 2 unique items}
\item{Elicit (\url{elicit.org}) 64 items, 61 unique}
\end{itemize}

\subsection{Paper Selection and Citation Snowballing}
For the 259 unique items we determined whether it is a peer-reviewed paper (no book, thesis or meta-research) that describes data engineering for AI. 

In total we excluded 240 items. With those items we performed card sorting on the reason for exclusion, resulting in eleven categories, see Table~\ref{tab:ExcludingItems}. The card sorting serves as a soundness check of the exclusion process (did we exclude for the right reason?). 

The selection process resulted in 19 (= 259 - 240) papers.  

\begin{table}
  \caption{Excluded items}
  \label{tab:ExcludingItems}
  \begin{tabular}{lc}
    \toprule
    &\textbf{\# Items}\\
    \midrule
    Book, thesis or meta-research & 52\\
    Not peer-reviewed	            & 53 \\
    No data engineering activities 	& 28 \\
    No AI engineering - Software engineering & 3 \\
    No AI engineering - Data Science &53 \\
    No AI engineering - AI governance   &10 \\
    No AI engineering - Education	&5  \\
    No AI engineering - Infrastructure, Hardware  &14 \\
    No AI engineering - AI4SE          &5 \\
    Only mentions, no explanation & 7 \\
    Only challenges, no solutions & 10 \\
     \midrule
    \textbf{Total excluded}	& \textbf{240}  \\
  \bottomrule
\end{tabular}
\end{table}

With the 40 papers that were initially included based on title and abstract, we also performed snowballing according to the guidelines provided by ~\cite{WohlinEASE2014}. We checked all the references in the 40 papers, but also checked all citations of these 40 papers with Google Scholar. We repeated this snowballing process until no new papers were added. The complete snowballing process has added 6 new papers to this final set, resulting in 25 papers in total (see Table \ref{Table_numbers}). 

\begin{table}
\centering
\begin{smaller}
\caption{Filtering publications on data engineering for AI}
\label{Table_numbers}
\begin{tabular}{l|l|l|l}
\toprule
& \textbf{Query} & \textbf{Inclusion} & \textbf{References} \\
\midrule
Google Scholar & 182 & 11 & 15 \\
ScienceDirect & 2 & 0 & 0\\
ACM DL & 1 & 0 & 0 \\
DBLP   & 13 & 1 & 1 \\
Elicit & 61 & 7 & 9\\
\midrule
\textbf{TOTAL} & \textbf{259} & \textbf{19} & \textbf{25} \\
\bottomrule
\end{tabular}
\end{smaller}
\end{table}

\section{Paper Classification}\label{sec:data}
The resulting set of 25 papers was classified according to different dimensions. This section explains the different classifications. 

\subsection{Classification by Meta-Data}
\label{sec_meta}
To indicate the background of the selected papers we classified them by the following meta-data.  
\begin{enumerate}
\item Author affiliation (Aff. = University, Company, Public Organization, Research Center)
\item Author country;
\item Year of publication;
\item Number of pages (\#p);
\item Number of citations (\#cit.);
\item Keywords;
\item Focus on software engineering (SE) or focus on data science (DS).
\end{enumerate}

\begin{figure}
  \centering
  \includegraphics[width=\linewidth]{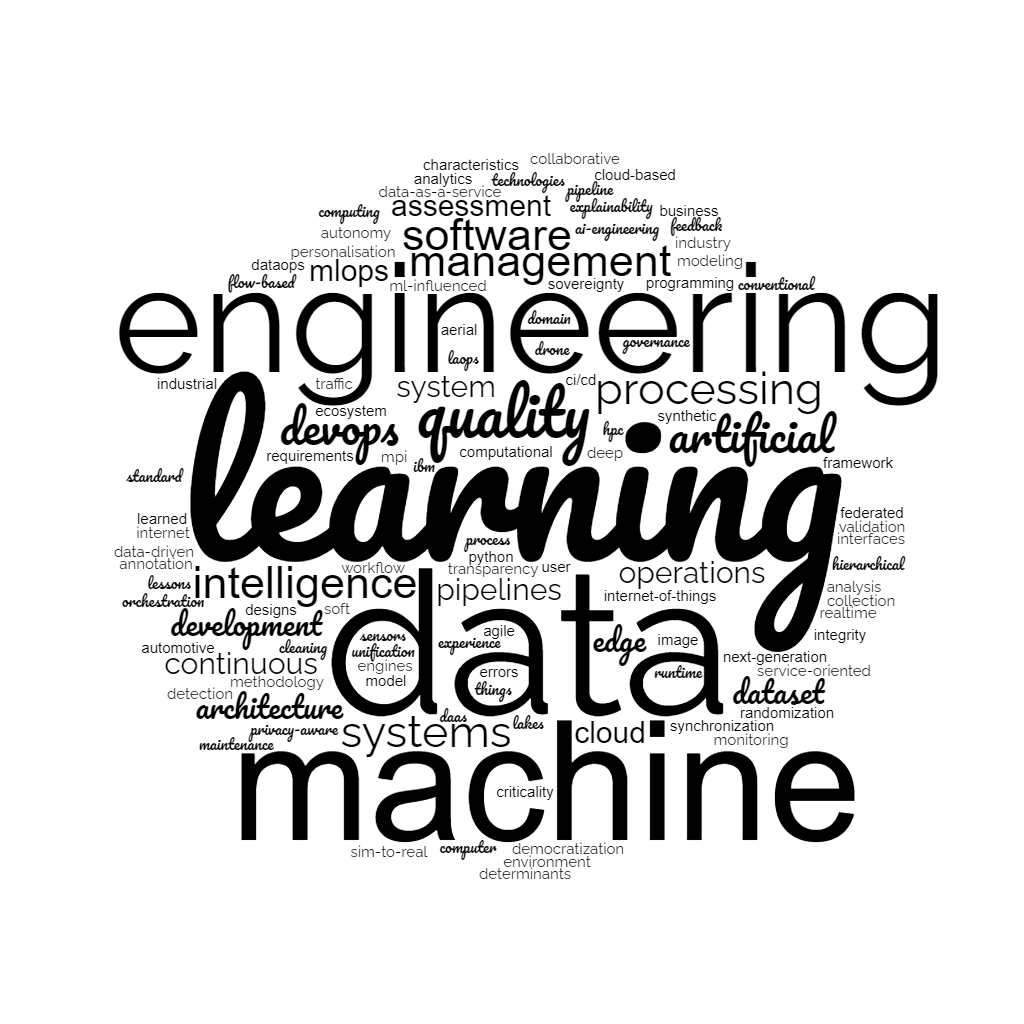}
  \caption{Word cloud of the keywords from the 25 selected papers}
  \Description{Word cloud of the keywords from the 25 selected papers}
  \label{fig:keywords}
\end{figure}

\subsection{Classification by Scope of Data Engineering}
To indicate the scope of data engineering activities explained in each paper we classified them according to the following categories.
\begin{enumerate}
\item Data pipeline for training AI systems (TP);
\item Data pipeline for serving data to AI systems in production (PP);
\item System-wide AI data architecture (DA);
\item Enterprise-wide AI data architecture (EDA);
\end{enumerate}

\subsection{Classification by Type of Data Engineering Solution}
To indicate what type of guidance on data engineering for AI the paper provides we classified them according to the following categories.  
\begin{enumerate}
\item Technical solution (tool, platform, library, etc.);
\item Architecture;
\item Best practices;
\item Case study;
\end{enumerate}

To provide better guidance for practitioners, we also indicate for each paper which of the following empirical validation of the provided solutions was done.
\begin{enumerate}
\item Case study in industry;
\item Validation within author's company;
\item Experiments;
\item Demo applications or implementations;
\item No empirical validation.
\end{enumerate}

\subsection{Classified Papers}
Table \ref{Table_classes} shows how each of the 25 selected papers classifies on metadata, scope, and solution type.

\begin{table*}[h]
\centering
\begin{smaller}
\caption{Papers on data engineering for AI}
\label{Table_classes}
\begin{tabular}{l l l|l l|l l|l|l|l|l}
\toprule
\textbf{ID}& \textbf{Ref.} &\textbf{Year} & \textbf{Affiliation} &\textbf{Country} &\textbf{\#p} & \textbf{\#cit.} & \textbf{SE/DS} & \textbf{Scope} & \textbf{Type} & \textbf{Validation}\\
\midrule
DE1&~\cite{breck} & 2019 & Company (Google) & USA & 12 & 238 & SE & PP & Technical solution & Case study\\
DE2&~\cite{derakhshan}&2019&University&Germany&12 & 23&DS& PP&Technical solution & Experiments\\
DE3&~\cite{frost}&2019&University&USA&4 & 4&SE&PP&Technical solution & Demo\\
DE4&~\cite{yokoyama}&2019&Company (Fujitsu)&Japan&8 & 41&SE&DA&Architecture& Demo\\
DE5&~\cite{abeykoon}&2020&University&USA, Sri Lanka&9 & 5&SE&TP&Technical solution & Experiments\\
DE6&~\cite{munappy2020ad}&2020&Univ. + Comp. (Ericsson)&Sweden&10 & 66&SE&EDA&Best practices & Company\\
DE7&~\cite{groger}&2021&Company (Robert Bosch)&Germany&9 & 36&SE&EDA&Architecture & Company\\
DE8&~\cite{Lawakatare}&2021&Univ. + Comp. (Ericsson)&Sweden&10 & 13&SE&PP&Best practices & Company\\
DE9&~\cite{raj2021impact}&2021&Univ. + Research C.&Sweden&10 & 1&SE&PP&Best practices & Case study\\
DE10&~\cite{Saravanan}&2021&Research Center&Qatar&9 & 2&DS&TP&Architecture & No\\
DE11&~\cite{Altendeiterung}&2022&Research C. + Univ.&Germany, Spain&12 &  0&SE&TP&Case study & Case study\\
DE12&~\cite{chattopadhyay}&2022&Univ. + Public Org.&USA&6 & 1&DS&TP&Case study & Case study\\
DE13&~\cite{Cheng}&2022&University&Australia&6 & 1&SE&TP&Best practices & No\\
DE14&~\cite{foidl}&2022&Univ. + Research C.&Austria&11 & 11&SE&PP&Technical solution & Experiment\\
DE15&~\cite{hasterok}&2022&Research Center&Germany&10 & 5&SE&DA&Best practices & No\\
DE16&~\cite{niemela}&2022&University&Finland&8 & 0&SE&DA&Case study & Case study\\
DE17&~\cite{paleyes}&2022&University&UK&11 & 3&SE&DA&Architecture & Demo\\
DE18&~\cite{petersen}&2022&Research Center&Germany&8 & 1&SE&TP&Best practices & Demo\\
DE19&~\cite{sen}&2022&Research Center&Norway, Spain&7 & 2&SE&EDA&Architecture & Case study\\
DE20&~\cite{wang}&2022&Research Center&China&6 & 1&DS&TP&Architecture & Demo\\
DE21&~\cite{warnett}&2022&University&Austria&11 & 4&SE&TP&Architecture & No\\
DE22&~\cite{azimi}&2023&University&Italy&19 & 0&DS&PP&Architecture & Demo\\
DE23&~\cite{jariwala}&2022&University&India, Vietnam&14 & 1&DS&TP&Technical solution & Demo\\
DE24&~\cite{kreuzberger}&2023&Univ. + Comp. (IBM)&Germany&10 & 49&SE&TP&Architecture & Interview\\
DE25&~\cite{sabet}&2023&Univ. + Comp. (Microsoft)&USA&17 & 0&DS&TP&Technical solution & Experiments\\
\bottomrule
\end{tabular}
\caption*{SE=software eng., DE=data science, TP/PP=training/production pipeline, DA/EDA=system/enterprise-wide data architecture}
\end{smaller}
\end{table*}

What stands out in this table is that the set of 25 papers comes from a mix of industry and academics, with only 9 being purely academic. Most papers (15 out of 25) come from European countries, with a large number from Germany (6) and Sweden (3). Note that the 3 papers from Sweden seem to come from the same research group. Most papers (17) focus on the software engineering perspective, which is not surprising, since we specifically selected on ``AI engineering''. With respect to the types of solutions discussed in the papers, all of them are well represented. Furthermore, most papers contain some form of empirical evaluation of their solutions. 

With respect to the scope of activities explained in the paper, eleven focus on the training pipeline (TP) and seven focus on the production pipeline (PP). Only four papers focus on system-level data engineering and only three on enterprise-level data engineering. This means most papers take quite a narrow definition of data engineering: setting up a data pipeline for a machine learning model.   

With respect to the keywords, the list is long and quite diverse, see the word cloud in Figure \ref{fig:keywords}. Papers DE1, DE2 and DE14 are without keywords, the other papers have selected between three and ten (DE16) keywords. DE13, DE15, DE16, DE17, DE21 and DE24 do not mention data in their keywords, but they do mention AI or machine learning. The other papers all selected keywords that include ``data'':
\begin{itemize}
    \item data processing: DE3, DE4
    \item data engineering: DE5, DE12
    \item DataOps: DE6
    \item data pipelines: DE3 (pipeline), DE6, DE9
    \item data technologies: DE6
    \item data management, data democratization, data governance, data ecosystem: DE7
    \item data quality: DE8, DE23
    \item data errors: DE8
    \item data validation: DE8
    \item data transparency: DE10
    \item data cleaning: DE10
    \item data sovereignty: DE11
    \item data-driven development: DE18
    \item data integrity: DE19
    \item hierarchical dataset, dataset design: DE20
    \item data collection standard, data synchronization: DE20
    \item Data-as-a-Service, DaaS: DE22
    \item synthetic data: DE25
\end{itemize}

This list of keywords already shows that terminology within the 25 papers is not standardized and that many different aspects of data engineering for AI are being considered. We will analyze this in more detail in the next section, related to the research questions. 

\section{Data Engineering for AI Systems}\label{RQ}
In this section we discuss our findings related to each of the four research questions. To answer the research questions, the full text of each paper was coded according to 1) the life cycle phases that are described; 2) the technical solutions that are described; 3) the architecture pictures it contains; 4) the lessons learned it contains.

\subsection{RQ1: Which Data and AI Engineering Life Cycle Phases Are Covered?}\label{sec:RQ1}
To map the 25 selected papers to life cycle phases, we coded them with the AI engineering life cycle phases from Figure \ref{fig:phasesAIEng} (DATA, ML, DEV and OPS) and the data engineering life cycle phases from Figure \ref{fig:DE} (Generation, Ingestion, Transformation, Serving). When the paper does not focus on one or more specific life cycle phase(s), we coded it with ``All''. Table \ref{Table_lifecycle} shows the division of the papers over the life cycle phases. 

\begin{table}
\centering
\begin{smaller}
\caption{RQ1: Which life cycle phases are covered?}
\label{Table_lifecycle}
\begin{tabular}{l|l|p{3.5cm}}
\toprule
\textbf{AI Eng.} & \textbf{Data Eng.} & \textbf{Papers} \\
\midrule
DATA & Generation & DE18, DE20, DE23, DE25\\
DATA & Transformation & DE10, DE12\\
DATA & All & DE6, DE21\\
DATA+ML & Transformation & DE3\\
DATA+ML+DEV & All & DE17\\
DEV & All & DE4\\
DEV+OPS & All & DE9, DE19\\
OPS & Serving & DE1, DE2, DE8, DE14, DE22\\
All & All & DE5, DE7, DE11, DE13, DE15, DE16, DE24\\
\bottomrule
\end{tabular}
\end{smaller}
\end{table}

\textbf{Conclusion}. Not surprisingly the majority of papers cover at least the DATA phase of the AI engineering life cycle. But eight of the papers focus more on the DEV and/or OPS part of the AI engineering life cycle. Out of them, five papers specifically focus on data validation in production (the Serving phase of the Data Engineering life cycle).

\subsection{RQ2: Which Technical Solutions for AI Data Engineering Are Proposed?}\label{sec:RQ2}

As can be seen in Table \ref{Table_classes}, seven of the papers discuss technical solutions for AI data engineering. In this section we discuss each of the seven proposed solutions in more detail. 

[DE1] Breck et al.~\cite{breck} present a ``data validation system that is designed to \textit{detect anomalies specifically in data fed into machine learning pipelines}. This system is deployed in production as an integral
part of TFX – an end-to-end machine learning platform at Google.'' They discuss the challenges they faced in developing the system and the techniques they used to address them, including design choices that were made. They also present three case studies at Google to illustrate the benefits of the data validation system in production. 

[DE2] Derakshan et al.~\cite{derakhshan} propose a ``\textit{platform for continuously training deployed machine learning models} and pipelines that adapts to the changes in the incoming data.'' Their platform uses techniques such as proactive training, online statistics computation and dynamic materialization to reduce (re)training and deployment costs. They include evidence from experiments, with two different machine learning pipelines.

[DE3] Frost et al. present ``AI Pro, an open-source framework for \textit{data processing} with Artificial Intelligence (AI) models.'' With AI Pro users can generate a data pipeline from a configuration file through a user friendly web interface. For advanced users and core developers, there is a command line interface for in-depth operations with finer-grained control. They demonstrate AI Pro with two demo scenarios.

[DE5] Abeykoon et al.~\cite{abeykoon} developed a ``\textit{high performance Python API} with a C++ core to represent data as a table and provide distributed data operations.'' Their PyCylon solution bridges ETL pipelines in Python (as mostly used by data scientists) with high performance compute kernels in C++. They conducted experiments to proof the performance of PyCylon.  

[DE14] Foidl et al.~\cite{foidl} collected a catalogue of 36 ``data smells'' in a multi-vocal literature review and implemented \textit{tool support to detect these data smells}. They applied the tools to 246 Kaggle datasets to evaluate them. As opposed to the data anomalies detected by the system of [DE1] presented above, these data smells are broader, since smells also include ``\textit{potential} data quality issues''.   

[DE23] Jariwala et al.~\cite{jariwala} demonstrate the use of the \textit{IBM Data Quality for AI Toolkit} to check training data in a machine learning setting. They include a workflow how to call the IBM API and show the results of several included data quality metrics on open source datasets.  

[DE25] Sabet et al.~\cite{sabet} introduce ``a scalable Aerial \textit{Synthetic Data Augmentation} (ASDA) framework tailored to aerial autonomy applications.'' They demonstrate the ASDA framework by generating data for landing pad detection in the Seattle simulation scene. Although this is a very specific technical solution, the usage of synthetic datasets is of course not limited to aerial autonomy applications.

\textbf{Conclusion}. The selected papers contain a diverse set of technical solutions, ranging from synthetic data generation (DE25), through data validation tools (DE1, DE14, DE23), through data processing frameworks (DE3, DE5), to deployment platforms (DE2). In that way the solutions presented together cover the complete AI engineering and data engineering life cycles, although most solutions focus on one single life cycle phase (see Table \ref{Table_lifecycle}). The exception is [DE5] that does not cover one specific life cycle phase and is the only paper that covers the DEV phase of AI engineering.  

\subsection{RQ3: Which Architectures for AI Data Engineering Are Proposed?}\label{sec:RQ3}

As can be seen in Table \ref{Table_classes}, nine of the papers discuss architectures for AI data engineering. In this section we discuss each of the nine architectures in more detail. 

[DE4] Yokoyama~\cite{yokoyama} propose a \textit{multi-layer architectural pattern for machine learning systems} that separates the business logic from the inference engine and data processing. Furthermore, it separates the user interface from the data collection and the data lake from the database. They demonstrate their architectural pattern by designing a chatbot system.

\begin{figure}[h]
  \centering
  \includegraphics[width=\linewidth]{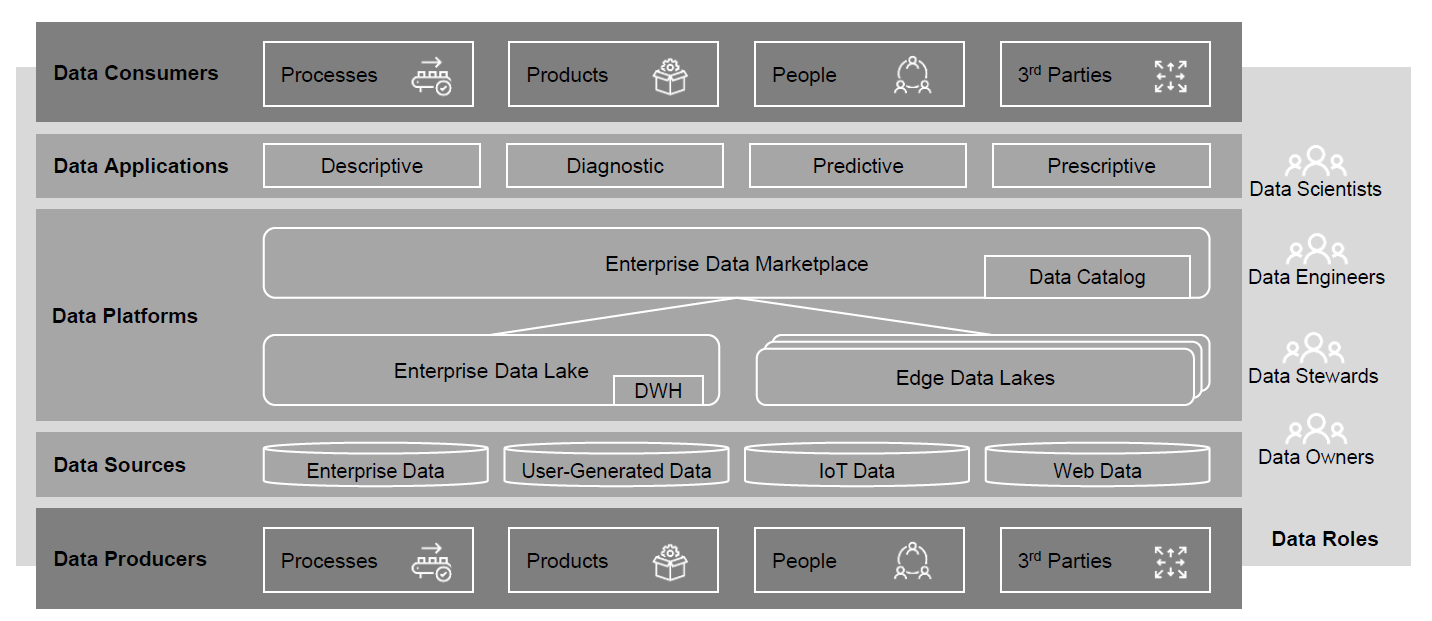}
  \caption{A data ecosystem for industrial enterprises~\cite{groger}}
  \Description{A data ecosystem for industrial enterprises~\cite{groger}}
  \label{fig:ecosystem}
\end{figure}

[DE7] Gr\"oger~\cite{groger} calls for a \textit{data ecosystem} for industrial enterprises, see Figure \ref{fig:ecosystem}. That ecosystem contains a specific role for data engineers and data engineering as part of the data democratization challenge: ``making all kinds of data available for AI for all kinds of end users across the entire enterprise''. Gr\"oger suggests to address this challenge with an enterprise data catalog that provides comprehensive metadata management across all data lakes and other data sources. This would enable self-service use of data.

[DE10] Thirumuruganathan et al.~\cite{Saravanan} present a \textit{reference architecture for automated annotations of data}. They describe the key components of this system architecture. Implementing a proof-of-concept remains future work. 

[DE17] Paleyes et al.~\cite{paleyes} propose ``\textit{Flow-Based Programming} as a paradigm for creating Data Oriented Architecture (DOA) applications.'' They compared the flow-based programming (FBP) paradigm to the Service-Oriented Paradigm (SOA) by implementing four data-driven applications in both paradigms and measuring evolution of the codebase through pre-defined metrics.  

[DE19] Sen et al.~\cite{sen} devised a ``\textit{de-centralized edge-to-cloud architecture}'' with machine learning pipelines for erroneous data repair and detection of deviations in sensor data. They analyze their proposed architecture in two different industrial case studies.  

[DE20] Wang et al.~\cite{wang} implement a \textit{hierarchical dataset} with unified annotation rules. They use one example scenario to compare a hierarchical dataset created from three single datasets with the original single-source dataset.

[DE21] Warnett and Zdun~\cite{warnett} list \textit{architectural design decisions (ADDs) for the machine learning workflow} from a gray literature study. Their replication package contains in an ADD model with UML diagrams of all ADDs and their relations. 

[DE22] Azimi and Pahl~\cite{azimi} present a \textit{layered Data-as-a-Service (DaaS) quality management architecture}. Their framework focuses on input data quality and links it to machine learned data service quality. They demonstrate their framework with a traffic management use case.

[DE24] Kreuzberger et al.~\cite{kreuzberger} depict an ``\textit{end-to-end MLOps architecture and workflow} with functional components and roles''. The workflow contains a separate data engineering zone and data(Ops) engineer is a separate role (apart from e.g., software engineer, data scientist or even ML engineer). According to them a data engineer ``builds up and manages data and feature engineering pipelines'' and ``ensures proper data ingestion to the
databases of the feature store system''. This indicates that their architecture/workflow has the scope of one single ML project.    

\textbf{Conclusion}. Most architectures presented focus on (parts of a) system architecture (DE4, DE10, DE17, DE20, DE22) or the ML pipeline (DE21, DE24). Only papers DE7 and DE19, contain a diagram for enterprise-wide data architectures, of which DE19 focuses on IoT data only. The data ecosystem from DE7 contains a comprehensive overview of this enterprise data landscape, including IoT data sources, see Figure \ref{fig:ecosystem}.

\subsection{RQ4: What are Lessons Learned on AI Data Engineering?}\label{sec:RQ4}

As can be seen in Table \ref{Table_classes}, nine of the papers discuss case studies or best practices for AI data engineering. In this section we discuss each of the nine papers in more detail.

[DE6] Raj et al.~\cite{munappy2020ad} derive a definition of DataOps from literature including the main components identified. According to their analysis, DataOps can be defined as ``an approach that accelerates the delivery of high quality results by automation and orchestration of data life cycle stages.'' They describe eight DataOps use cases at Ericsson and derived a five stage \textit{DataOps evolution} from them. For each stage they define requirements, which can also be read as best practices.   

[DE8] Lwakatare et al.~\cite{Lwakatare} conducted action research at Ericsson for \textit{training data validation}. Based on their research, they propose a data validation framework that considers multiple levels of data validation checks (feature, dataset, cross-dataset, data stream) and provision of feedback. Their research also identified three best practices: 1) define data quality tests at all four levels, 2) provide actionable feedback and suggest mitigation strategies, 3) treat data errors with similar rigor as code. 

[DE9] Raj et al.~\cite{raj2021impact} conducted a \textit{multiple-case study into six data pipelines} from commercial software-intensive systems at three companies. They describe and explain seven determinants for data pipelines: 1) Big Data, 2) data preprocessing, 3) data quality, 4) data storage requirements, 5) data pipeline elements, 6) performance efficiency, 7) continuous monitoring and fault detection. These determinants are important factors to consider when implementing a data pipeline. 

[DE11] Altendeiterung et al.~\cite{Altendeiterung} conducted a twelve-month \textit{action research at Mondragon on data-sovereign AI pipelines}. They explicitly list ten lessons learned: 1) need for data traceability, 2) need for an independent trustee, 3) need for quality-driven data sharing, 4) need for a data catalog, 5) need for real-time support, 6) need for a separation of control and data plane, 7) need for access and usage control enforcement, 8) need for standardization, 9) need for a common definition of user roles, 10) need for a trusted and secure deployment environment.

[DE12] Chattopadhyay et al.~\cite{chattopadhyay} describe an experiment where they used six different ways to \textit{transform a rainfall dataset from a five-minute span to a single value} (baseline, mean, median, mode, maximum, minimum). They call this choice a ``data engineering decision'' and argue that the choice made impacts model results quantitatively and qualitatively. 

[DE13] Cheng and Long~\cite{Cheng} propose \textit{Federated Learning Operations (FLOps) as a methodology for developing cross-silo federated learning systems}. They describe a life cycle with three phases and fourteen activities (one of which is data engineering). They also provide three best practices: 1) metadata engineering to create Data Interface Abstractions (DIAs) for models, 2) dual deployment of models and DIAs, 3) check points between phases. 

[DE15] Hasterok and Stompe~\cite{hasterok} present the \textit{Process Model for AI Systems Engineering (PAISE\textsuperscript{\textregistered})} as an extension of ISO/IEC 15288 with AI engineering. PAISE\textsuperscript{\textregistered} specifies its own procedures for ML component development and data provisioning. The data provisioning process facilitates the development of datasets as separate components in the system. It also is a feedback loop, where datasets can be updated based on the monitoring of the system.   

[DE16] Niemel\"{a} et al.~\cite{niemela} propose an \textit{architecture for a learning analytics system (LAOps)} in which they combine MLOps and privacy-aware cryptographic data storage. They describe the LAOps implementation they currently have at Tampere University.   

[DE18] Petersen et al.~\cite{petersen} introduce a \textit{data-driven workflow for developing qualitative datasets in automative systems engineering}. They showcase this process by ``curating a data pool consisting of different available data sources that have to be
integrated to cover as many driving situations as possible.''

\textbf{Conclusion}. The lessons learned from the above nine papers are quite diverse and with different scope. DE6 about DataOps is probably the most broad one together with DE15 about PAISE\textsuperscript{\textregistered} (process model for AI engineering), whereas DE12 describes just one data wrangling step in one single project. Depending on the context of the AI engineering project, different papers apply: DE11 and DE13 for federated learning, DE16 for learning analytics and DE18 for automotive systems engineering. DE8 and DE9 focus on one specific part of the life cycle: automated training data validation (DE8) and data pipelines for serving data to machine learning models in production (DE9).  

\section{Discussion}\label{sec_discuss}
This section discusses the findings in light of the overall research question ``\textit{How to do data engineering for AI systems?}''

\subsection{Threats to Validity}
Most threats to validity in such a mapping study relate to researcher bias in selecting and coding papers. We mitigated this by 1) following the guidelines suggested by Kitchenham and Charters~\cite{Kitchenham}, Sald\~{a}na~\cite{Saldana2011} and Wohlin~\cite{WohlinEASE2014}; 2) documenting and reviewing all steps we made; 3) using existing life cycle models and definitions for the coding; 4) making available the entire dataset, including selection and coding for other researchers to validate our results. 

Note that because we specifically searched for ``AI engineering'' in the query string, we might have missed papers that refer to AI engineering with other wording. We mitigated this by also searching with AI-based tool Elicit (61 new papers of which 7 were included in the result set) as well as by snowballing from the other selected papers.  

\subsection{Defining Data Engineering for AI Systems}
Before we can answer the question ``How to do AI data engineering?'', we must first answer the question ``What is AI data engineering?''. All 25 selected papers take a different angle. Data engineering might refer to one single task or step in an AI engineering project, a discipline within software engineering or data science, or an enterprise-wide competency, see also Table \ref{Table_numbers} (TP, PP, DA or EDA). 

Out of all papers, DE12 takes the most narrow view on data engineering. Chattopadhyaya et al.~\cite{chattopadhyay} write about data engineering decisions for AI-based applications. They use this rather umbrella term to indicate the decision how they should convert a timestamp dataset into an interval-based dataset: take the mean, the mode, the maximum or the minimum for that interval.

Only 4 out of 25 papers actually define what they mean by data engineering. Paper DE5 by Abeykoon et al.~\cite{abeykoon} defines data engineering as ``The complex process of transforming raw data to a form suitable for analytics''. Paper DE6 by Raj et al.~\cite{munappy2020ad} has a rather fuzzy definition of data engineering as a step that ``performs two
different operations at the high level, which include data collection and data ingestion''. Paper DE7 by Gr\"oger~\cite{groger} defines data engineering as ``modelling, integrating and cleansing of data.'' Paper DE13 by Cheng and Long~\cite{Cheng} says ``the raw data in each entity is extracted, transformed, and prepared for model training''. These definitions are all also much more narrow than the definitions from Andrew Ng and Reis and Housley as given in the introduction. 

To get a complete picture of data engineering for AI systems, one should  combine the life cycles in Figure \ref{fig:phasesAIEng} and Figure \ref{fig:DE}. In that way, one has both a picture of how AI engineering connects to data as well as how data engineering connects to AI. We did not find any comprehensive work that already does this and only three papers that take a similar enterprise-level view on data engineering (DE6, DE7 and DE19).  
\subsection{Implications for Practitioners}
The mapping of the selected papers to life cycle phases (RQ1, see section \ref{sec:RQ1}) and type of solutions they provide (RQ2 till RQ4, see section \ref{sec:RQ2} till \ref{sec:RQ4}) provides guidance to practitioners which solutions to select for which project or activity. Running our mapping study, we also had the following observations on AI data engineering that could be useful for practitioners.
\paragraph{Big Data} During snowballing, we excluded a number of papers on data engineering for Big Data, that did not have an explicit reference to AI systems. However, those papers might contain valuable solutions for both researchers and practitioners that also holds for AI systems, as these systems are mostly Big Data systems as well. That kind of analysis was out of scope for this paper and might be a topic for a future mapping study: ``How to do data engineering for Big Data?''
\paragraph{Data quality} A number of papers relate to data quality or data validation. There might be an interesting body of knowledge (and tools) on those topics that was out of scope for our mapping study. Zhang et al.~\cite{zhang} list several methods for data testing in their survey on machine learning testing. The concept of data smells~\cite{foidl, shome} is also important to consider in this context as these indicate data quality issues that might lead to machine learning problems, different from data errors (see DE14~\cite{foidl}). 
\paragraph{Grey literature} In our mapping study we included only peer-reviewed papers. However, we found a number of other interesting resources: 1) The original post of Figure \ref{fig:phasesAIEng}~\cite{Farah} that already defined the DATA cycle as in fact being a DataOps process, 2) the blogs and whitepapers on DataOps referenced in DE6~\cite{munappy2020ad}, 3) the book ``Fundamentals of Data Engineering''~\cite{Reis} from which we borrowed Figure \ref{fig:DE}, 4) other books such as ``Data Fabric and Data Mesh Approaches with AI''~\cite{hechler} and ``Practical DataOps: Delivering agile data science at scale''~\cite{atwal}. This means that practitioners should definitely consider sources from grey literature on DataOps and modern data architectures (such as data meshes and data fabrics). 
\paragraph{Open source tooling} The open source tooling landscape for data is becoming bigger and bigger, see also the online blog post ``The State of Data Engineering''\footnote{\url{https://lakefs.io/blog/the-state-of-data-engineering-2023/}}. These kind of tools are necessary to achieve higher levels of DataOps, see DE6~\cite{munappy2020ad}.  
\paragraph{Data spaces} Some AI data engineering projects require data sharing between different organizations or entities. To support federated learning and data sovereignty several technical solutions such as Gaia-X, FIWARE and the International Data Space have been built up. Paper DE11 by Altendeiterung et al.~\cite{Altendeiterung} investigates how to integrate such solutions with AI pipelines, but we recommend to also keep an eye on the evolution of the data space solutions, since they are fairly new. 
\paragraph{Domain-specific data engineering} Two domains that stood out in our mapping study are data engineering for Internet-of-Things (IoT) and data engineering for automotive. There might be more literature or guidance on AI data engineering if one dives into a specific domain. 
\paragraph{Synthetic data} Paper DE25 by Sabet et al.~\cite{sabet} describes synthetic data generation. That is a topic that might not be relevant for all AI engineering projects, but if it is, we would like to point out that there is a whole body of knowledge (and tools) about synthetic data generation specifically that can be looked into by practitioners.

\subsection{Implications for Researchers}
The mapping of the selected papers to life cycle phases (RQ1, see section \ref{sec:RQ1}) and type of solutions they provide (RQ2 till RQ4, see section \ref{sec:RQ2} till \ref{sec:RQ4}) provides guidance to researchers to see what is already there and what is still missing. Running our mapping study, we also had the following observations on AI data engineering that could be useful for researchers.
\paragraph{Data engineering} Tebernum et al.~\cite{tebernum2021derm} developed a ``data engineering reference model (DERM) which outlines the important building blocks for handling data along the data lifecycle.'' They aim to bridge between data engineers and software engineers by providing a common ground for engineering data-intensive applications. They view data engineering as a sub-discipline of data science (``preparing data for data scientists''). The AI engineering research community could benefit from integrating with the data science research community on the data engineering topic. The DERM presented by Tebernum et al. could serve as a common ground also for this purpose. Tebernum et al. also show ample opportunities for future data engineering research.   
\paragraph{DataOps} Paper DE6 by Raj et al.~\cite{munappy2020ad} points to DataOps as an overall process to automate and orchestrate data life cycle stages. The large amount of references to grey literature they used indicates that DataOps is not receiving enough attention in research yet. In line with the evolution from DevOps to MLOps, there is also a need to evolve DataOps as a separate research field. This means that researchers should not consider data engineering as a single activity in a machine learning project, but as an approach that accelerates the entire enterprise data life cycle. 
\paragraph{Data-Oriented Architecture (DOA)} Paper DE17 by Paleyes et al.~\cite{paleyes} introduces the term DOA as opposed to Service-Oriented Architecture (SOA). DOA seems an interesting paradigm for AI data engineering, but a quick search in Google Scholar only yields 23 articles that relate to DOA. More research is needed to establish if and how DOAs can be used to solve AI engineering challenges. 
\paragraph{Enterprise-wide data architectures} The previous section points practitioners to books about data fabrics and data meshes. These are new concepts for managing data within an enterprise. The question how to effectively engineer AI systems making use of these concepts, remains open. 
\paragraph{Data spaces and federated learning} The same holds for data spaces and federated learning. Paper DE11 by Altendeiterung et al.~\cite{Altendeiterung} investigates how to combine IDS with AI pipelines, but more research is needed on combining data spaces with federated learning.
\paragraph{Production data} Most papers focus on data engineering for training data. Now that AI has matured and more and more projects go into production, we need more background on how to engineer production data pipelines, how to validate and monitor production data, and how to set up enterprise-wide production data architectures. 
\paragraph{Open source tooling} Open source tooling could also be a vehicle for researchers to transfer results to practitioners. In the area of AI engineering, open source tooling is wide-spread in industry, so researchers can easily integrate their data engineering solutions. 
\paragraph{Knowledge engineering} Mattioli et al.~\cite{mattioli} contrasts data-driven AI and knowledge-driven AI and argue that a hybrid approach is needed to build trustworthy AI systems. They point to the discipline of knowledge engineering, separate from data engineering. According to them, ``knowledge engineering (KE) is the process of understanding
and then representing human knowledge in data structures, semantic models (conceptual diagram of the data as it relates to the real world) and heuristics.'' In that way, it complements data engineering as it creates the data structures that data engineering deals with. This paper specifically focuses on data-driven AI engineering, but it is an interesting question how to combine this with knowledge-driven AI engineering. 
\paragraph{Systems engineering} We focused our mapping study on AI engineering, thinking it to be a discipline within software engineering. However, DE15 about PAISE\textsuperscript{\textregistered} and DE18 about a data-driven workflow describe AI engineering as a discipline within systems engineering. Then the data engineering part is not about enterprise-wide data, but about data within one system (e.g., device or machine). The AI engineering research community could benefit from integrating with the AI system engineering research community, as they might run into similar data-related challenges.   

\section{Conclusion}
In this paper we created an overview of existing literature on data engineering for AI systems from an AI engineering perspective. 

We found that most papers focus on engineering training or production pipelines for AI systems, but that they lack overall data architecture guidance for AI systems or the AI-driven enterprise. For software engineers and software engineering researchers this means that after DevOps and MLOps, now DataOps (and the integration between the three) is a new important topic to address. There is a strong need for frameworks, best practices, but also open source tools to support practitioners in implementing them. This paper provides a first overview of what is already there. 

Future work remains to update the analysis, preferably also include available grey literature and books, and learn from case studies what is missing in practice. Our ultimate goal is to develop a data engineering toolbox for AI engineers, that includes both tooling to support project-level data pipelines as well as enterprise level data architectures. And, most importantly, an integrated data engineering and AI engineering approach.     

\begin{acks}
This research has been co-financed by ``Regieorgaan SIA'', part of the ``Nederlandse Organisatie voor Wetenschappelijk Onderzoek (NWO)'' and Fontys Kenniscentrum Applied AI for Society.
\end{acks}

\bibliographystyle{ACM-Reference-Format}
\bibliography{References}

\end{document}